\documentclass[aps,prl,reprint,preprintnumbers,showpacs,showkeys,superscriptaddress,nofootinbib,floatfix]{revtex4-1}
\usepackage{graphicx}
\usepackage{epsfig,amssymb,amsmath,color}
\usepackage{times}
\usepackage[colorlinks=true,citecolor=blue,linkcolor=blue]{hyperref}

\usepackage{ulem}
\usepackage{subfigure}

\newcommand{\beq}{\begin{equation}}
\newcommand{\eeq}{\end{equation}}
\newcommand{\bea}{\begin{eqnarray}}
\newcommand{\eea}{\end{eqnarray}}
\newcommand{\bear}{\begin{array}}
\newcommand {\eear}{\end{array}}
\newcommand{\bef}{\begin{figure}}
\newcommand {\eef}{\end{figure}}
\newcommand{\bec}{\begin{center}}
\newcommand {\eec}{\end{center}}
\newcommand{\non}{\nonumber}

\newcommand{\la}{\left\langle}
\newcommand{\ra}{\right\rangle}

\begin{document}
\title{Disappearing Inflaton Potential via Heavy Field Dynamics}

\author{Naoya Kitajima}
\email{kitajima@tuhep.phys.tohoku.ac.jp}
\affiliation{Department of Physics, Tohoku University, Sendai 980-8578, Japan}

\author{Fuminobu Takahashi}
\email{fumi@tuhep.phys.tohoku.ac.jp}
\affiliation{Department of Physics, Tohoku University, Sendai 980-8578, Japan}
\affiliation{Kavli Institute for the Physics and Mathematics of the Universe (WPI), 
UTIAS, University of Tokyo, Kashiwa 277-8583, Japan
}

\begin{abstract}
We propose a possibility that the inflaton potential is significantly modified
after inflation due to heavy field dynamics.  During inflation there may be a heavy scalar field
stabilized at a value deviated from the low-energy minimum. As the heavy field moves to the
low-energy minimum, the inflaton potential could be significantly modified. In  extreme cases, 
the inflaton potential vanishes and the inflaton becomes
almost massless at some time after inflation. Such transition of the inflaton potential has interesting 
implications for primordial density perturbations, reheating, creation of unwanted relics, dark radiation,
and experimental search for light degrees of freedom.
To be concrete, we consider a chaotic inflation in supergravity where the inflaton mass parameter is
promoted to a modulus field, finding that the inflaton becomes stable after the transition and contributes 
to dark matter.
Another example is the new inflation by the MSSM Higgs field which acquires a large expectation value
just after inflation, but it returns to the origin after the transition and settles down at the electroweak vacuum.
Interestingly, the smallness of the electroweak scale compared to the Planck scale
is directly related to the flatness of the inflaton potential.

\end{abstract}
\preprint{TU-1005, IPMU15-0144}
\maketitle

\paragraph{Introduction}
-- The inflation \cite{Guth:1980zm,Kazanas:1980tx,Starobinsky:1980te,Sato:1980yn,Brout:1977ix,Linde:1981mu,Albrecht:1982wi} 
is strongly supported by observations of the temperature and
polarization anisotropies of cosmic microwave background (CMB) \cite{Ade:2015xua,Ade:2015lrj}.
While there have been proposed various inflation models, yet it is unclear what the
inflaton is, how the reheating occurs, etc. In particular,  it is considered difficult to pin down the inflation model
because so far the inflaton sector can be probed only through gravitational interactions.

In many inflation models, the inflaton has
a heavy mass (e.g. at intermediate scales), which is well beyond the reach of any foreseen particle collider 
experiments.\footnote{There are various exceptions such as
the Higgs inflation with a non-minimal coupling to gravity~\cite{Bezrukov:2007ep} 
or a running kinetic term~\cite{Nakayama:2010sk}. 
} 
This is partly because the inflaton mass is usually determined by
some combination of the normalization of CMB power spectrum 
of order $10^{-5}$ and the Planck mass, unless the  model contains 
additional small parameters or other dimensionful parameters.
The above argument, however, is based on the constancy of the inflaton mass, or more precisely, 
of the inflaton potential after inflation. This may not be the case if there are other dynamical degrees of freedom,
which evolve significantly after inflation.

There appear many moduli fields in supergravity/string theory through compactifications
of  the extra dimensions. Those moduli fields must be stabilized for successful low-energy
effective theory. While many of them can be stabilized with heavy masses by fluxes, 
some may remain relatively light and play an important role in cosmology.
For instance, suppose that there is a modulus field which is stabilized only by supersymmetry (SUSY)
breaking effects. Such modulus field generically acquires  a mass of order the Hubble parameter during inflation,
and its position could be largely deviated from the low-energy minimum. During or after inflation, the modulus field
may evolve with time toward the low-energy minimum, which  affects the inflaton potential.  If this happens during inflation,
the prediction of primordial density perturbations is modified \cite{Rubin:2001in,Dong:2010in,Achucarro:2010da,Cespedes:2012hu,Harigaya:2015pea}.
If this happens after inflation, on the other hand, the inflaton potential, and therefore the properties of the inflaton
 can be significantly modified.

In this Letter we study a possibility that the inflaton potential is significantly modified
by heavy field (modulus) dynamics. In particular we focus on the case in which the inflaton potential vanishes and
the inflaton becomes almost massless after inflation. As we shall see, this occurs when some
symmetry, which is spontaneously broken during inflation, becomes restored after inflation.
Such transition  has interesting 
implications for primordial density perturbations, reheating, creation of unwanted relics, dark radiation,
and  experimental search for light degrees of freedom.
For instance, as the inflaton mass becomes significantly light,  non-thermal production
of gravitinos~\cite{Kawasaki:2006gs,Asaka:2006bv,Dine:2006ii,Endo:2006tf,
Endo:2006qk,Endo:2007ih,Endo:2007sz} can be suppressed. 
Some massless (or light) degrees of freedom may appear at the enhanced symmetry  point (ESP),
contributing to dark matter or dark radiation. 
Also, the inflaton must have couplings with the standard model (SM)
particles for successful reheating. Therefore, the experimental search for the inflaton may become
possible. For instance, the inflaton shows up as an axion-like particle, if the inflaton 
enjoys a shift symmetry. To be concrete we will provide a couple of inflation models where such
transition takes place.

\vspace{1mm}

\paragraph{Disappearing Inflaton Potential}
\label{sec2}
-- 
In order to illustrate our main idea,  let us consider the following Lagrangian, 
\beq
	\mathcal{L} = \frac{1}{2}K(s)\partial_\mu \phi \partial^\mu \phi + \frac{1}{2} G(s) \partial_\mu s \partial^\mu s - V(\phi,s),
\eeq
where $\phi$ and $s$ are the inflaton and a modulus field, respectively.
For simplicity we do not consider the dependence of the kinetic term coefficients on $\phi$,
which would lead to the so called running kinetic inflation~\cite{Takahashi:2010ky,Nakayama:2010kt}.
In our scenario, the inflaton potential is modified by non-trivial evolution of the modulus field $s$
after inflation.
To this end, we consider the potential with
\beq
\label{Vphi}
	V(\phi,s) = F(s) v(\phi)+U(s).
\eeq
Such a form of the potential is adopted for illustration purposes, and a more general form of the 
potential is possible. 
For a given value of $\phi$, $F(s)$ and $U(s)$ determine the potential minima of $s$.

Let us suppose that, during inflation,  $s$ is stabilized at  one of the local minima, $s=s_{\rm inf}$, 
which is largely deviated  from the low-energy minimum, $s = 0$. At $s=s_{\rm inf}$,
$F(s)$ takes a nonzero value so as to drive the inflation with the effective potential,
\begin{align}
V_{\rm eff}(\phi) =  F(s_{\rm inf}(\phi))  v(\phi) +  U(s_{\rm inf}(\phi)), \label{eff_potential}
\end{align}
where $s_{\rm inf}$ may depend on $\phi$, which changes the effective inflaton potential.\footnote{A classical example is the smooth hybrid inflation model~\cite{Lazarides:1995vr}.}
After inflation, as the energy density of the inflaton decreases,
the $s$  moves to the global minimum.
In particular, $s$ may be attracted to the ESP where some symmetry  is restored,
and massless degrees of freedom may appear~\cite{Kofman:2004yc}.
In extreme cases we have $F(0) = U(0)=0$, and the inflaton potential disappears.
Thus, the inflaton becomes almost massless after inflation.

Similarly, the kinetic term coefficients $K(s)$ and $G(s)$ can be modified significantly during/after inflation, and as a result,
the inflaton mass and couplings are modified after field redefinition to satisfy the canonical normalization. 
Integrating out the $s$ field during inflation, such inflaton dynamics  can be modeled 
by the running kinetic inflation~\cite{Takahashi:2010ky,Nakayama:2010kt}. If the transition occurs after inflation,
it can lead to stronger couplings between the inflaton and the SM particles in the present vacuum,
 while much weaker couplings can be realized in order not to spoil the slow-roll inflaton dynamics during inflation.

\vspace{1mm}
\paragraph{Chaotic Inflation Models in Supergravity}
\label{sec3}
-- 
Here we provide a concrete inflation model where the inflaton potential
disappears after inflation.
As a simple example, let us consider an extension of the chaotic inflation 
model in supergravity~\cite{Kawasaki:2000yn} (see also \cite{Kallosh:2010ug}). In our model, the inflaton mass parameter 
in the original model is promoted to a chiral superfield $S$. We will see that
the properties of the inflaton $\Phi$ crucially depends on
 the dynamics of $S$.

In order to assure the trans-Planckian excursion of the inflaton field, we impose a shift symmetry, $\Phi \to \Phi +iC$,
where $C$ is a real transformation parameter. 
We assume that the K\"ahler potential 
respects the shift symmetry at tree level;
\beq
	\begin{split}
	K &= \frac{1}{2} (\Phi + \Phi^\dag)^2(1+c_\Phi^{(2)} |S|^2 + c_\Phi^{(4)} |S|^4 ) \\
	& + |X|^2 (1+c_X^{(2)} |S|^2 + c_{X}^{(4)} |S|^4) \\
	& + |S|^2 (1+c_S^{(2)} |S|^2 + c_S^{(4)}|S|^4 ) + \cdots,
	\end{split}
\eeq
where  $X$ is a stabilizer field,
$c_\Phi^{(i)}$, $c_X^{(i)}$ and $c_S^{(i)}$ are 
 constants of order unity and the dots represent irrelevant or higher order terms suppressed by the Planck mass.
 Here and in what follows, we adopt the Planck units in which the reduced Planck mass is set to be unity, 
 $M_P = 1$. We introduce a symmetry breaking term in the superpotential,
\beq
	W = \lambda S X   \Phi,
\eeq
 which lifts the inflaton potential to drive inflation. Here
 $\lambda$ is a dimensionless coupling constant, and it can be regarded as an order parameter for the symmetry breaking.
 This model has ${\rm U(1)_R}$, U(1)$_S$ and ${\rm Z_2}$ symmetries
 with the charge assignment, $\Phi(0,0,-1)$, $X(2,1,1)$, and $S(0,-1,-1)$.\footnote{
{Note that the charge assignment is not unique. }The U(1)$_{R,S}$ symmetries may be replaced with their discrete
 subgroups, if one would like to add other interactions of $S$ to lift the scalar potential.
 The $Z_2$ symmetry is imposed for simplicity. Our main result holds in models without $Z_2$.
 } 
 If $S$ develops a nonzero vacuum expectation value (VEV) during inflation,
U(1)$_S$ and $Z_2$ are spontaneously broken. If $S$ is stabilized at the origin after inflation, those symmetries are restored.
 
\begin{figure}[tp]
\centering
\includegraphics [width = 7.0cm, clip]{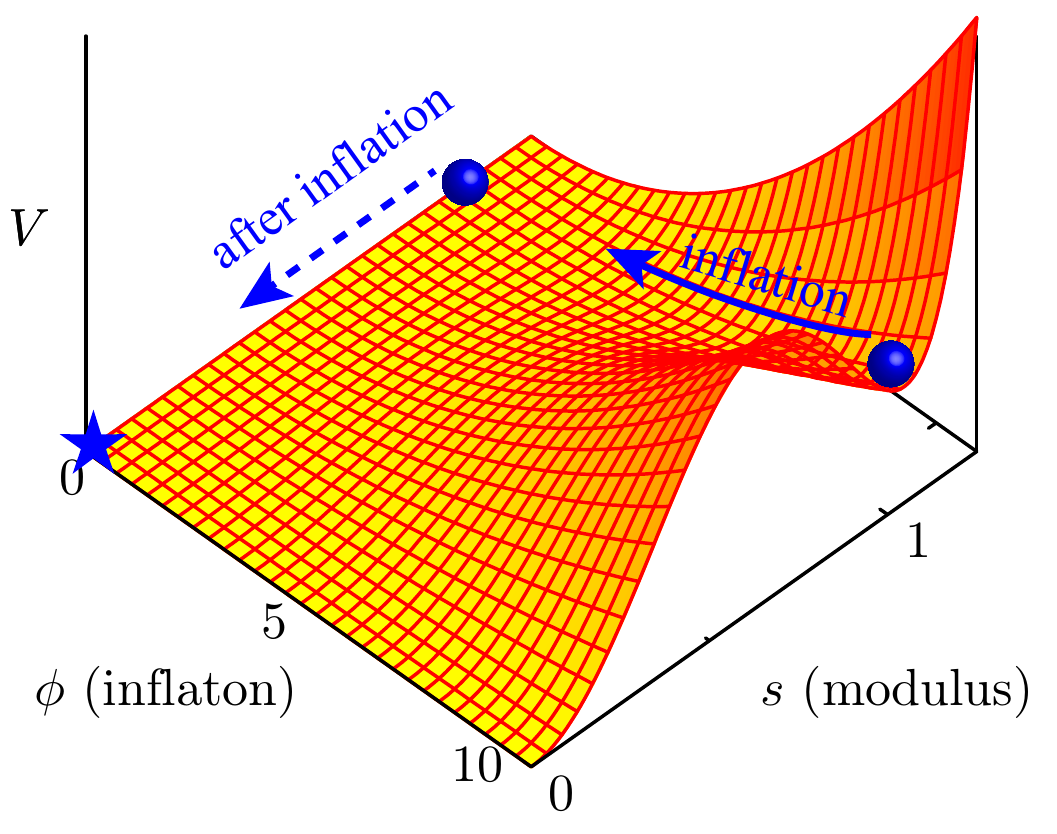}
\caption{
Schematic view of the potential (\ref{vphi}).
}
\label{fig:potential}
\end{figure}

The scalar potential in supergravity is given by
\bea
	V &=& e^K \bigg[ (D_i W) K^{i{\bar j}}  (D_j W)^\dag - 3|W|^2 \bigg],
\eea
where $D_i W = \partial_i W + K_i W$.
Let us decompose each scalar component (denoted by the same character as that of the superfield) as 
$\Phi = (\eta+i\phi)/\sqrt{2}$,  and $S = s e^{i\theta_s}/\sqrt{2}$. Then $\phi$ plays a role of the inflaton, 
and $s$ is the modulus field.  We assume that, during inflation,  the stabilizer field $X$ as well as  $\eta$
are stabilized at the origin with a mass of order the Hubble parameter. This is possible if there
 are operators like $|X|^2 (\Phi+\Phi^\dag)^2$ and $|X|^4$ in the K\"ahler potential.
Then 
the kinetic term for $\phi$ and $s$ and the scalar potential are given by
\beq
	\begin{split}
	\label{LK}
	{\cal L}_K=&\frac{1}{2} \partial_\mu \phi \partial^\mu \phi \bigg( 1+\frac{1}{2} c_\Phi^{(2)} s^2 +\frac{1}{4}c_\Phi^{(4)} s^4 + \dots \bigg) \\
	&+ \frac{1}{2} \partial_\mu s \partial^\mu s \bigg(1 + 2c_S^{(2)} s^2 +\frac{9}{4}c_S^{(4)}s^4 +\dots \bigg),
	\end{split}
\eeq
\begin{align}
	V(\phi, s) &=e^K K^{X \bar X} |W_X|^2,\\
	&= \frac{1}{4}\lambda^2 s^2 \phi^2 \bigg( 1 - \frac{1}{2} c_2 s^2 + \frac{1}{4} c_4 s^4 + \dots \bigg).
	\label{vphi}
\end{align}
The effective potential takes the form of Eq.~(\ref{Vphi}) with
\begin{align}
F(s) &= \frac{1}{4}\lambda^2 s^2 \bigg( 1 - \frac{1}{2} c_2 s^2 + \frac{1}{4} c_4 s^4 + \dots \bigg),\non\\
\label{Veff}
v(\phi)&=\phi^2,~~~U(s)  = 0,
\end{align}
where we define $c_2 = c_X^{(2)}-1$ and $c_4 = (1-2c_X^{(2)}(1-c_X^{(2)})-2c_X^{(4)}+2c_S^{(2)})/2$. 
In order for successful inflation, $s$ must be stabilized at a nonzero value, $s_{\rm inf} \ne 0$,
 where the potential has a positive definite vacuum energy.
These conditions, i.e., $\langle s_{\rm inf} \rangle \neq 0$ and $V(\langle s_{\rm inf} \rangle) > 0$, are translated to 
$c_2 > 0$ and $c_2^2/4 < c_4 < c_2^2/3$, if we neglect the higher order terms.
Then $s$ can be stabilized near the Planck scale
with a mass of order the Hubble parameter. See Fig.~\ref{fig:potential} for the scalar potential.
Note that, since $U(s)$ vanishes, $s_{\rm inf}$ does not depend on the inflaton field value, 
leading to the simple quadratic inflaton potential. The inflation takes place along $\phi$ 
since all the other degrees of freedom are fixed with a heavy mass of order the Hubble parameter.

Some time after inflation $s$  moves to the origin, as  the global minimum is located at $S=X=\Phi=0$,  where 
the inflaton potential disappears and the inflaton as well as other degrees of freedom become (almost) massless.
Once $s$ starts to move, the inflaton becomes much lighter, which affects the reheating process if it has not
completed yet.
Taking account of the SUSY breaking effects,  all the fields except for the inflaton $\phi$ acquire
masses of order the gravitino mass, $m_{3/2}$, through Planck-suppressed couplings with the SUSY breaking sector.
  The inflaton mass is
protected by the shift symmetry, and so, its mass is suppressed by the order parameter for the symmetry 
breaking; 
\beq
\label{infmass}
m_{\phi,0}^2 \;\sim\; \frac{\lambda^2}{16 \pi^2} m_{3/2}^2,
\eeq
where the subscript $0$ indicates the value at the present epoch.

A couple of comments are in order. The quadratic chaotic inflation is strongly disfavored by the current 
CMB observations~\cite{Ade:2015lrj}, which necessitates some extension. One simple extension is to consider a polynomial
chaotic inflation model~\cite{Nakayama:2013txa,Nakayama:2013jka,Kallosh:2014xwa,Nakayama:2014wpa}, and 
the superpotential is extended to
\beq
W = \lambda S X \Phi  \left(1+ k_2 \Phi + k_3 \Phi^2+ \cdots \right),
\label{polyW}
\eeq
where $k_{2,3}$ are numerical coefficients and the dots represent higher order terms. 
The second term is suppressed in the presence of $Z_2$ symmetry.
The predicted values of the spectral index $n_s$ and the tensor-to-scalar ratio $r$ can be well within the
allowed range if $|k_2|$ or $|k_3|^{1/2} = {\cal O}(0.1)$. 
This is because one needs to make the inflaton potential flatter around 
$\phi \sim {\cal O}(10)$
to reduce the value of $r$.

The Planck normalization of the density perturbations is satisfied if $F(s_{\rm inf}) \sim 10^{-10}$.
Barring cancellations, this  requires $\lambda \sim 10^{-5}$, because
$s_{\rm inf}$ is close to the Planck scale.  This implies that
the shift symmetry is of high quality, as $\lambda$ is regarded as an order parameter for the shift 
symmetry breaking. On the other hand, it may be possible
that the local minimum at $s = s_{\rm inf}$ is quasi-degenerate with the global one at $s = 0$.\footnote{
Topological inflation~\cite{Vilenkin:1994pv,Linde:1994hy} may occur
at a saddle point, where $s$ plays a role of the inflaton before the polynomial 
chaotic inflation due to $\phi$. This may explain the initial condition of the chaotic inflation
with $\lambda \sim 0.1$.
}
Then, the Planck normalization is satisfied for a larger value of $\lambda$, e.g. $\lambda \sim 0.1$.
This implies that the shift symmetry is not of high quality, and the potential along $\phi$ may become
steep at $\phi \gtrsim {\cal O}(10)$~\cite{Nakayama:2014hga}.
Interestingly, this is consistent with the polynomial
 chaotic inflation (\ref{polyW}) where the higher order terms are accompanied with numerical
 factors $|k_2|$ or $ |k_3|^{1/2} = {\cal O}(0.1)$. 
The smallness of the density perturbation could be partly due to the quasi-degenerate local minimum along the
heavy degrees of freedom, rather than due to small parameters in the inflation model.

In the above set-up, the heavy field $s$ is stabilized at $s_{\rm inf}$, which is constant during inflation.
It is possible to introduce additional terms like $S^n$ in the superpotential, which leads to
a nonzero $U(s)$. As a result,  $s_{\rm inf}$ evolves with time during inflation, 
which affects $n_s$ and $r$ (see e.g. \cite{Harigaya:2015pea}).

\paragraph{Reheating and cosmological implications}
\label{sec4}
-- 
Now let us study the evolution of the Universe after inflation.
After $\phi$ starts to oscillate around the origin, the (local) potential minimum of $s$ is generically deviated from
$s_{\rm inf}$, as the kinetic energy of $\phi$ modifies the potential (cf. Eq.~(\ref{LK})). Depending on the couplings,
$s$ will gradually move to the origin soon after inflation or stay around the Planck scale until later times.
In the former case, the inflaton becomes much lighter soon after inflation, and both the inflaton and $s$ field continue to
oscillate, interacting with each other. Effectively their energy density decreases like radiation until the soft SUSY breaking
mass terms dominate the potential. Finally $s$ decays into light degrees of freedom, and the inflaton will be
dark matter, whose abundance likely exceed the observed dark matter abundance unless there is a late-time
entropy production. To avoid the overclosure of the Universe, 
let us here focus on the latter case in which $s$ stays around
the Planck scale until its soft SUSY breaking mass becomes relevant after reheating.

The inflaton can decay into right-handed neutrinos through the superpotential,
\beq
	W = y \Phi S N N + \gamma \Pi N N, \label{PhiNN}
\eeq
where $y$ and $\gamma$ are coupling constants, and $\Pi$ is a chiral superfield whose VEV
 generates the right-handed neutrino mass~\footnote{
If one consider the superpotential~\cite{Kawasaki:2000ws},
\beq
	W = y \Phi N N + \gamma \Pi N N,
\eeq
the inflaton gets a large VEV, generating a large SUSY mass for $S$ and $X$.
}, and we have suppressed the flavor indices.
The charge assignment is summarized in Table \ref{tab:charge}, where we replaced U(1)$_S$ with
Z$_{4B-L}$.
The coupling $y$ is expected to be of order $\lambda$ as it breaks the shift symmetry.
When $\Pi$ develops a nonzero VEV, Z$_{4B-L}$ gets broken to Z$_{2B-L}$. 
Note that the $Z_2$ symmetry keeps the inflaton from acquiring a large VEV at that time and $s$ 
remains stabilized at the false vacuum until the inflaton decays.\footnote{
The following superpotential
\beq
	W = m_{3/2}S^2 + \lambda \langle \Pi \rangle m_{3/2} S \Phi
\eeq
can also be allowed but it does not affect our analysis.
}
The decay rate of the inflaton into the lightest right-handed neutrinos is estimated as
\beq
	\Gamma_\phi \sim \frac{(y s_{\rm reh})^2}{16 \pi} m_{\phi} \sim 20~{\rm GeV} \bigg(\frac{y s_{\rm reh}}{10^{-5}} \bigg)^2 \bigg( \frac{m_{\phi}}{10^{13}~{\rm GeV}} \bigg),
\eeq
where $s_{\rm reh}$ denotes the value of $s$ at the inflaton decay. The reheating temperature is of order $10^9$\,GeV or higher.
Then thermal or non-thermal leptogenesis~\cite{Fukugita:1986hr} works successfully unless there is a huge late time entropy production.

\begin{table}[tb]
\begin{center} {\tabcolsep = 2mm
	\begin{tabular}{c|ccccccc} \hline
		\rule[0mm]{0mm}{4.0mm} & $\Phi$ & $S$ & $X$ & $N$ & $\Pi$ & $Q$ & $\bar{Q}$ \\ \hline
		U(1)$_R$ & 0 & 0 & 2 & 1 & 0 & 1 & 1\\
		Z$_2$ & $-$ & $-$  & $+$ & $+$ & $+$ & $-$ & $+$ \\ 
		Z$_{4B-L}$ & 0 & 2  & 2 & 1 & 2 & 1 & 1 \\ \hline
	\end{tabular} }
\end{center}
\caption{Charge assignment}
\label{tab:charge}
\end{table}

When the soft mass of $s$, $m_{s,0}$, becomes comparable to the Hubble parameter,
it starts to oscillate about the origin and the inflaton as well as $X$ become much lighter.
In the case discussed above, the inflaton has already decayed at the commencement of oscillations 
of the $s$ field. For this,  $y \simeq 10^{-4}$ is required for $m_{s,0} \lesssim 1$ TeV,
if the inflaton mainly decays via (\ref{PhiNN}).
As the initial oscillation amplitude is about the Planck scale,
 coherent oscillations of $s$ soon dominates the energy density of the Universe.
The decay of $s$ therefore must thermalize the SM sector.
To this end let us introduce the superpotential $W = h S Q\bar{Q}$, 
where $Q$ and $\bar{Q}$ are hidden quarks 
having 
a hidden gauge symmetry and U(1)$_R$, Z$_2$ and Z$_{4B-L}$ charges listed in Table~\ref{tab:charge}.
The decay of $s$ into the hidden quarks occurs relatively soon after the commencement of 
oscillation for $h = \mathcal{O}(1)$. 
The produced hidden quarks will thermalize the SM sector, if there are  
massive hidden quarks in the bifundamental representation of the hidden and SM gauge groups,
or if the massive hidden quarks are charged with a hidden U(1)$_H$ gauge symmetry with a
kinetic mixing with U(1)$_Y$.
Note that the additional massless degrees of freedom are naturally expected in this case as a consequence of the ESP.
In addition, they can contribute to the dark radiation in the present Universe and may be detected by future observations.

Let us point out an interesting possibility.
After $s$ finally stabilized at the origin, the inflaton becomes stable due to the unbroken Z$_2$ symmetry.
Thus the inflaton can contribute to the present dark matter abundance.
Such inflaton dark matter can be produced as a relic of the incomplete reheating if $s$ starts to oscillate 
soon after the reheating, while the detailed calculation of the abundance of the inflaton dark matter is left for  future work.

If the Z$_2$ symmetry is slightly broken, 
the inflaton may also be  coupled to photons (or other SM gauge fields) through 
\beq
	\mathcal{L} = \xi \frac{\phi }{M_P} F \tilde{F}, \label{phiFFtilde}
\eeq
where $\xi$ is an order parameter for the Z$_2$ breaking.\footnote{
Note that, if the kinetic term coefficient in the present vacuum
is sufficiently suppressed, $K(0) \ll 1$, the effective coupling is enhanced by a factor 
of $1/\sqrt{K(0)}$.} Then, the inflaton may behave like a decaying dark matter like axion like particles.
Note however that, once the $Z_2$ symmetry is explicitly broken,  extra Z$_2$ breaking terms 
can be added to the superpotential and they may alter the above scenario.
For instance, the term like $\xi SX$ in the superpotential is allowed, and the inflaton  acquires the VEV of 
order $|\xi / \lambda|$ which leads to a heavy mass of $S$ and $X$.\footnote{ 
This can be avoided if $\xi/\lambda$ is a pure imaginary value because such parameter can be absorbed 
by the inflaton field due to the shift symmetry.
}
In this case, $s$ decays into the hidden sector at early times and the inflaton is left as a stable cold dark matter.
Therefore subsequent entropy production is required to dilute the inflaton to be comparable to (or less than) the
observed dark matter abundance.

In case that Z$_2$ is largely broken, the inflaton can decay soon after the end of inflation through e.g. a coupling 
with the Higgs sector. The relevant superpotential is
\beq
	W = \kappa \Phi H_u H_d,
	\label{PHH}
\eeq
where $\kappa$ denotes the numerical coefficient of order $\lambda$ or smaller.
The inflaton decays into a pair of higgsino through the above interaction. 
The inflaton is finally stabilized at large field values, giving rise to a large mass for $S$ and $X$ and 
a large contribution to the $\mu$-term.\footnote{
Alternatively, one can consider $W = \kappa S \Phi H_u H_d$, where the $Z_2$
is spontaneously broken by the VEV of $S$, and the Higgs fields are assumed to
be charged under U(1)$_S$.  In this case there is no large contribution to the
$\mu$-term.
}
So, $s$ must start to oscillate and decay at early times, since otherwise, $s$ is kept stabilized near
the Planck scale and never rolls down to the origin. To this end one can include  non-minimal couplings
in the K\"ahler potential to generate $\dot{\phi}^2s^2$ in the Lagrangian.
This effectively behaves like a mass of $s$ and can drive $s$ to the origin.
Furthermore, the inflatino could be the lightest SUSY particle, contributing to dark matter.
The inflatino can be thermally produced through interactions with the SM particles, e.g. (\ref{PHH}).

\paragraph{More general inflation models}
--
So far, we have considered the minimal extension of the chaotic inflation model in 
which the mass parameter is promoted to the $S$ field.
In fact, the above stabilization of the modulus $S$ and the subsequent disappearing inflaton potential 
can be generalized to a broader class of inflation models. To this end, let us consider
the superpotential of the form,
\beq
W = X {\cal F}(\Phi,S)
\label{WXFS}
\eeq
where $X$ has a U(1)$_R$ charge $2$, and it is assumed to be stabilized at the origin
due to the quartic coupling in the K\"ahler potential. Then the scalar potential reads
\beq
V = e^K K^{X \bar X} \left|{\cal F}(\Phi,S) \right|^2,
\eeq
in the approximation of $\langle W  \rangle \approx 0$. For certain couplings, $S$ can be stabilized
by balancing terms in the first two prefactors. We may focus on a specific class of models with
${\cal F}(\Phi,S) = g(S) f(\Phi)$, leading to
\beq
	V =  e^K K^{X \bar X} \left|g(S) f(\Phi)\right|^2.
\eeq
For successful inflation, one needs to assume that $S$ is deviated from any of SUSY vacua determined by $g(S)=0$.\footnote{
This is possible if one chooses such $g(S)$ that the SUSY vacua other than $S=0$ are located at super-Planckian values.
}  For instance, one may take $g(S) = S^n$. During inflation $f(\Phi) \ne 0$, 
the $S$ can be stabilized by the K\"ahler potential around the Planck scale during inflation.
After inflation, $f(\Phi) = 0$,  the $S$ moves to the low-energy minimum by the soft SUSY breaking mass.

As an example, let us consider an extension of the new inflation model \cite{Asaka:1999yd,Asaka:1999jb,Senoguz:2004ky,Nakayama:2012dw},
\beq
	g(S) = S,~~f(\Phi) = -\mu^2 + \frac{\Phi^{2m}}{M_*^{2m-2}} 
\eeq
where $m~(\geq 2)$ is an integer value and $\mu$ and $M_*$ are respectively the energy scale of inflation and the cutoff scale.
In this model Z$_{2m}$ symmetry is imposed to ensure the flatness of the inflaton potential.
The scalar potential is calculated as
\beq
	V = \frac{v^4 s^2}{2} \bigg(1- \frac{2\phi^{2m}}{M^{2m}} + \frac{\phi^{4m}}{M^{4m}} \bigg) \bigg(  1 - \frac{c_2 s^2}{2} + \frac{c_4 s^4}{4} - \frac{\beta \phi^2}{2} + \dots \bigg).
\eeq
where we have defined $M = 2(v M_*^{m-1})^{1/m}$ which is the potential minimum of the inflaton for $\langle s \rangle \ne 0$ and 
dots in the second parenthesis represent the higher order Planck suppressed terms.
$X$ is assumed to be stabilized at the origin due to the positive Hubble mass during inflation.
$\beta$ is an $\mathcal{O}(1)$ coefficient determined by the non-minimal K\"ahler potential $|\Phi|^2 |X|^2$ and we assume $\beta >0$.
One can approximately decompose the potential in the form of (\ref{eff_potential}) with
\beq
\begin{split}
	F(s) &= \frac{1}{2} v^4 s^2 \bigg( 1 - \frac{1}{2}c_2 s^2 + \frac{1}{4} c_4 s^4 +\dots \bigg), \\[1mm]
	v(\phi) &= 1- \frac{\beta \phi^2}{2} -\frac{2\phi^{2m}}{M^{2m}} + \frac{\phi^{4m}}{M^{4m}} +\dots,\\
	U(s) &= 0.
\end{split}
\eeq
Similar to the case with the chaotic inflation model, $s$ is stabilized near the Planck scale by $F(s)$ during inflation while inflation takes place by $v(\phi)$.
After inflation, the inflaton rolls down to the potential minimum at $\phi =M$. 
The $s$ starts to roll down to the low-energy potential minimum at the origin 
when its soft mass term becomes significant.  Then the inflaton potential vanishes, and the inflaton becomes (almost) massless.

In contrast to the original model~\cite{Asaka:1999yd,Asaka:1999jb,Senoguz:2004ky},   the inflaton  eventually falls into another vacuum 
determined by the soft SUSY breaking effects. For instance, the inflaton may be stabilized at the origin, even though it acquired  a large VEV just after inflation.
This enables us to identify the inflaton field with one of the flat directions in SUSY SM such as $H_uH_d$.
The relevant superpotential in this case can be written as
\beq
	W = XS \bigg(-v^2 + \frac{(H_u H_d)^2}{M_*^2} \bigg) + \mu H_u H_d.
\eeq
The advantage of this scenario is that the reheating is automatic.
Note that this model is based on the Z$_4$ symmetry,
which is softly broken by the $\mu$-term. Therefore, the smallness of the electroweak scale compared to the Planck scale
is a natural outcome of  the discrete symmetry which  ensures the flatness of the inflaton potential. 
The hierarchy problem is connected to the flatness of the inflaton potential.

So far we have treated the modulus $S$ as a gauge singlet, but it may be similarly identified with one of the
flat directions in SUSY SM; e.g. $S^2 = H_uH_d$, $S^3 = udd, eLL$, and $S^4=QQQL$, etc., if we consider the superpotential
of the form $W = X S^n f(\Phi)$. It is also possible to identify both the inflaton $\Phi$ and the modulus $S$ with
flat directions which are compatible with each other. The reheating into the SM particles becomes automatic, and
the baryon asymmetry may be generated by the Affleck-Dine mechanism~\cite{Affleck:1984fy,Dine:1995kz}.

\paragraph{Conclusions}
\label{sec5}
--
In summary, we have proposed a possibility that the inflaton potential is significantly affected by the heavy field dynamics.
As an extreme case we have studied a chaotic inflation model in supergravity where the mass parameter is
promoted to a superfield $S$, and the inflaton becomes almost massless in the present vacuum, $\la S \ra = 0$.
The Universe is reheated by decays of the inflaton and $S$ fields. Interestingly, the inflaton and/or $S$ fields
may be searched for at various experiments because their masses are at most of order the gravitino mass,
much lighter than the typical inflaton mass in other models. 
The strength of the couplings to the SM particles may be enhanced if the kinetic term of the inflaton is 
affected by the dynamics of $S$. In general, there appear massless (or light) degrees of freedom at
the ESP. For instance, if the inflaton potential arises from the strong gauge dynamics, the corresponding gauge interactions
may become weakly coupled (asymptotically non-free) at the ESP,  where there appear many massless hidden quarks.
Such quarks as well as hidden gauge bosons may contribute  to self-interacting dark radiation~\cite{Jeong:2013eza}.
Thus, the heavy field dynamics may enable us to probe the inflaton sector at various experiments/observations
through interactions like (\ref{PHH}) and (\ref{phiFFtilde}). We have also shown that the transition can be applied
to a broader class of inflation models (cf. discussion below (\ref{WXFS})), and in particular, the inflaton as well as 
the modulus field can be identified with flat directions in SUSY SM.

\acknowledgments

This work is supported by MEXT Grant-in-Aid for Scientific research 
on Innovative Areas (No.15H05889 (F.T.) and No. 23104008 (N.K. and F.T.)), 
Scientific Research (A) No. 26247042 and (B) No. 26287039 (F.T.), 
and Young Scientists (B) (No. 24740135 (F.T.)),
and World Premier International Research Center Initiative (WPI Initiative), MEXT, Japan (F.T.).

\bibliography{dspinf}
\end{document}